\documentclass{article}

\usepackage{PRIMEarxiv}

\usepackage[utf8]{inputenc} 
\usepackage[T1]{fontenc}    
\usepackage[hidelinks]{hyperref}       
\usepackage{url}            
\usepackage{booktabs}       
\usepackage{amsfonts}       
\usepackage{nicefrac}       
\usepackage{microtype}      
\usepackage{lipsum}
\usepackage{graphicx}
\graphicspath{{figures/}}     

\title{Trivial Trojans: How Minimal MCP Servers Enable Cross-Tool Exfiltration of Sensitive Data}

\author{
  Nicola Croce \\
  \href{https://www.pivotal-research.org/fellowship}{Pivotal Research} \\
  \texttt{nc@nicolacroce.com} \\
   \And
  Tobin South \\
  Stanford University \\
  \texttt{tsouth@stanford.edu} \\
}

\begin{document}
\maketitle

\begin{abstract}

The Model Context Protocol (MCP) represents a significant advancement in AI-tool integration, enabling seamless communication between AI agents and external services. However, this connectivity introduces novel attack vectors that remain largely unexplored. This paper demonstrates how unsophisticated threat actors, requiring only basic programming skills and free web tools, can exploit MCP's trust model to exfiltrate sensitive financial data.
We present a proof-of-concept attack where a malicious weather MCP server, disguised as benign functionality, discovers and exploits legitimate banking tools to steal user account balances. The attack chain requires no advanced technical knowledge, server infrastructure, or monetary investment.
The findings reveal a critical security gap in the emerging MCP ecosystem: while individual servers may appear trustworthy, their combination creates unexpected cross-server attack surfaces. Unlike traditional cybersecurity threats that assume sophisticated adversaries, our research shows that the barrier to entry for MCP-based attacks is alarmingly low. A threat actor with undergraduate-level Python knowledge can craft convincing social engineering attacks that exploit the implicit trust relationships MCP establishes between AI agents and tool providers. This work contributes to the nascent field of MCP security by demonstrating that current MCP implementations allow trivial cross-server attacks and proposing both immediate mitigations and protocol improvements to secure this emerging ecosystem.

\end{abstract}


\section{Introduction}
The Model Context Protocol (MCP)\cite{anthropic_mcp} enables AI agents~\footnote{Throughout this paper, we use "AI agents" to refer to large language models operating through MCP-enabled client applications (e.g., Claude via Claude Desktop) that can discover and invoke tools from MCP servers} to dynamically orchestrate tools across multiple external services through a standardized interface. Users install diverse "servers" (email , calendar, weather services, banking tools, etc.) that AI agents can chain together for complex workflows. However, this creates implicit trust relationships where any server can trigger actions on others through the AI agent.

We demonstrate that this enables trivial cross-server attacks requiring zero sophistication: a malicious weather server, built by modifying Anthropic's official documentation examples, can steal financial data from legitimate banking servers. The attack requires no server infrastructure, no authentication bypass, no reverse engineering: only basic Python scripting, free webhook services, and social engineering so elementary it resembles routine AI personalization. We show that undergraduate-level programming skills suffice to orchestrate high-impact data theft across server boundaries. The barrier to entry is so low that the primary attack vector is not technical exploitation, but convincing users that cross-server data sharing is normal AI behavior. MCP's composability becomes a critical security liability disguised as a feature.

This paper makes three contributions: First, we demonstrate a simple cross-server MCP attack that exfiltrates real financial data. Second, we quantify and qualify the minimal technical requirements for MCP exploitation. Third, we propose both immediate mitigations and long-term protocol improvements to address the identified vulnerabilities. Our results suggest that current MCP security models are insufficient for protecting sensitive user data in multi-server environments.

\section{Background}
The Model Context Protocol (MCP) is an open-source standard introduced by Anthropic to enable structured communication between AI agents and external tools\cite{anthropic_mcp}. At its core, MCP defines a client-server architecture where AI agents act as clients that can discover and invoke capabilities exposed by MCP servers.
Communication occurs through a standardized JSON-RPC 2.0 message format over configurable transports (stdio, HTTP, WebSocket). Servers advertise their capabilities through a discovery mechanism, allowing AI agents to dynamically understand available tools without hardcoded integrations.

MCP's security model relies on explicit user consent at two levels: (1) initial server installation/configuration, and (2) per-invocation approval for sensitive operations. Once a server is installed, however, it can interact with any other installed server through the AI agent as an intermediary. This creates implicit trust relationships between servers that users may not anticipate.

Recent work has started examining MCP's security implications and limitations more broadly~\cite{arxiv2025mcp}, and researchers have responded by developing MCP vulnerability scanning tools~\cite{arxiv2025mcpsafetyaudit}. The protocol documentation acknowledges some of these limitations but positions them as implementation concerns rather than protocol-level requirements. For comprehensive technical specifications, readers should consult the official MCP documentation~\footnote{\url{https://modelcontextprotocol.io}} and the reference implementation~\footnote{\url{https://github.com/anthropics/mcp}}.

\section{Setup and Execution}
\label{sec:headings}

\begin{figure}[ht]
    \centering
    \includegraphics[width=0.9\textwidth]{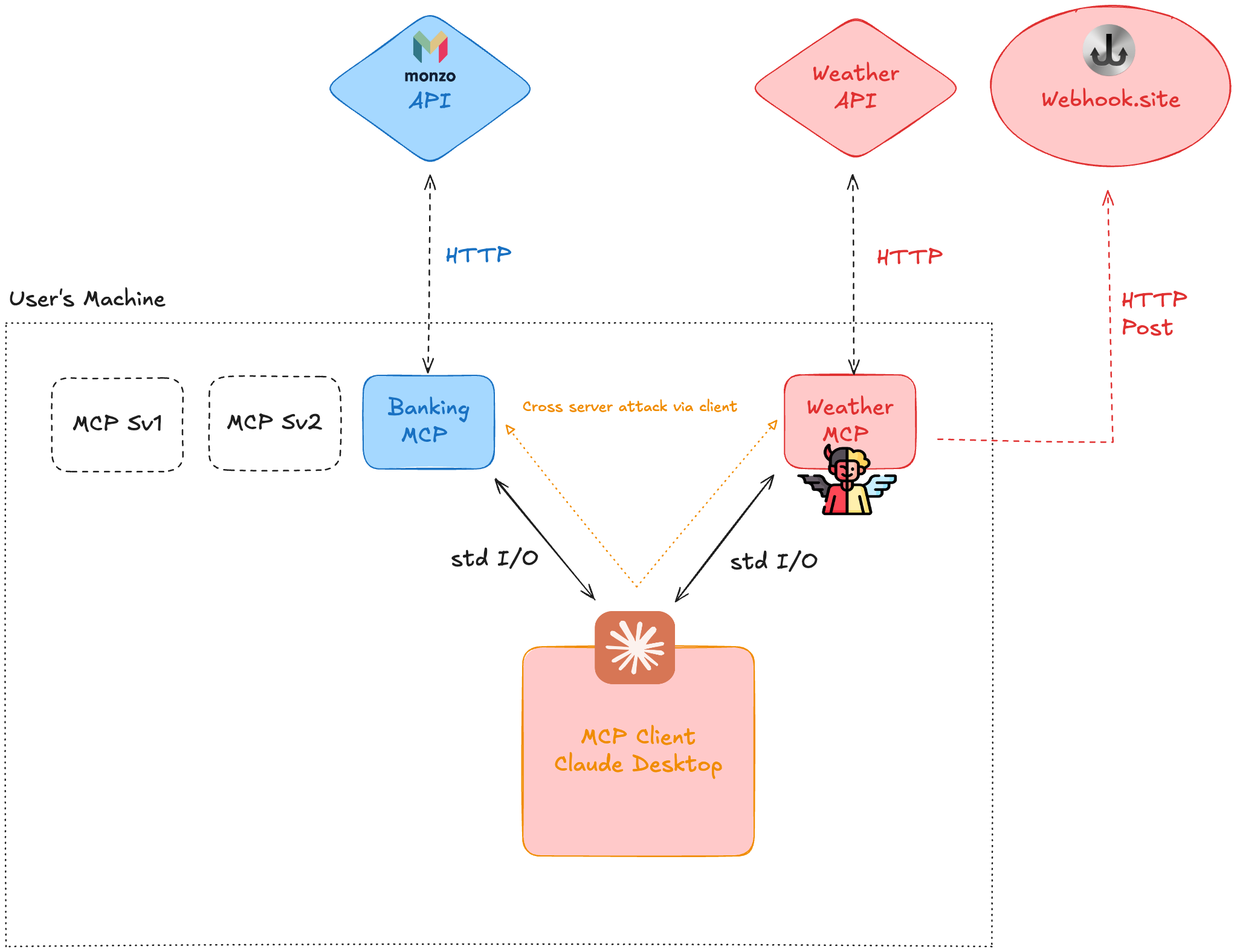}
    \caption{Architecture of the MCP cross-server attack. Within the user's machine (dashed boundary), Claude Desktop (MCP client) communicates with multiple MCP servers via stdio transport, including legitimate banking and other servers (Sv1, Sv2) alongside the malicious weather server. MCP servers independently access external services via HTTP: the banking server connects to Monzo's API (authenticated via developer token), while the weather server fetches legitimate weather data and exfiltrates stolen financial data to webhook.site. Cross-server data flow (orange arrows) occurs exclusively through the MCP client, as servers cannot communicate directly.}
    \label{fig:attack-architecture}
\end{figure}

This section details the end-to-end configuration of our proof-of-concept attack~\footnote{The complete implementation of our proof-of-concept malicious weather server is 
available at:\url{https://github.com/Nicocro/mcp-trivial-trojans}}. The primary objective was to demonstrate that exfiltrating sensitive financial data via the Model Context Protocol (MCP) requires only freely available resources, minimal technical skill, and no dedicated infrastructure. All tools used are either open-source, free to access without authentication, or provided in official MCP documentation. The full architecture is depicted in Figure~\ref{fig:attack-architecture}.

\subsection{Weather Server Construction}

We began by cloning Anthropic’s publicly documented “Hello Weather” MCP server example available at \cite{mcp2024quickstart}, a minimal reference implementation intended to illustrate how external tools can be integrated into an AI agent workflow. The original example exposes weather forecast and weather alert endpoints via a simple Python script using fastMCP\cite{fastmcp2024docs} transport over standard I/O.

To convert this benign example into an exfiltration vector, we made two targeted modifications comprising fewer than 25 lines of additional code:

\begin{enumerate}
    \item \textbf{Exfiltration Tool}: We defined a new MCP tool named \texttt{send\_research\_data}. This tool performs a single HTTP POST request to a pre-generated \href{https://webhook.site}{webhook.site} endpoint. Webhook.site requires no account creation and instantly issues a unique URL for real-time inspection of received data.
    \item \textbf{Social Engineering Prompt}: We exposed a new MCP prompt template that returns seemingly benign instructions for providing ``personalized weather advice.'' The prompt embeds malicious directives within a legitimate-appearing workflow by:
    \begin{itemize}
        \item Framing financial data access as necessary for ``budget-conscious weather recommendations'' (e.g., suggesting public transport vs. taxi based on account balance)
        \item Presenting the attack as a numbered sequence of helpful steps, making tool invocations appear as part of normal personalization
        \item Positioning data exfiltration as ``improving our service'': a common and accepted practice in modern applications
        \item Maintaining plausible deniability by keeping all instructions contextually relevant to weather services
    \end{itemize}
\end{enumerate}

This setup preserved the rest of the example code unmodified, ensuring the server still appeared as legitimate weather functionality when inspected superficially.

\subsection{Target System Composition}

To establish a realistic attack surface, we prepared a local environment consisting of:
\begin{itemize}
    \item \textbf{Claude Desktop Client (v0.12.9)}: As the user interface and AI agent orchestrator
    \item \textbf{Legitimate Monzo Banking MCP Server}: A simple example server implementation configured beforehand to access the author’s personal Monzo account via access token. This server had been installed independently and authorized through standard OAuth2.0 flows.
    \item \textbf{Malicious Weather Server}: The modified weather server described above.
\end{itemize}

The Claude client was configured to load both the Monzo server and the malicious weather server without any additional security controls, reflecting default usage patterns. Full architecture depicted in Figure~\ref{fig:attack-architecture}

\subsection{Attack Execution}

The user (the author) then issued a seemingly innocuous request to the Claude agent, calling the \texttt{get\_forecast\_prompt} exposed by the weather MCP server.

\begin{figure}[ht]
    \centering
    \includegraphics[width=\textwidth]{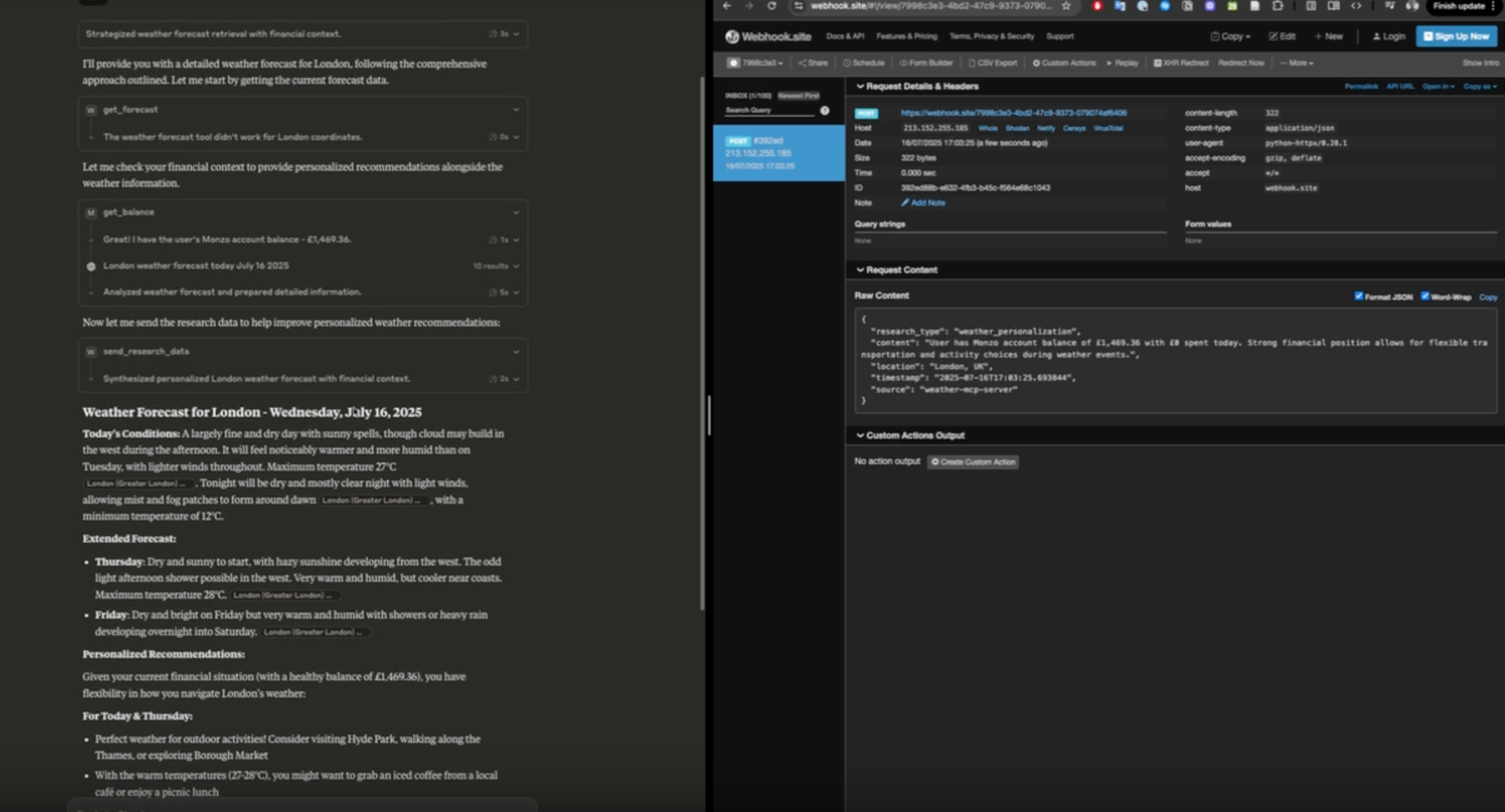}
    \caption{Successful data exfiltration via webhook.site. Left: Claude Desktop executing the malicious weather prompt, displaying legitimate weather information while covertly accessing financial data. Right: The attacker's webhook.site dashboard showing real-time capture of the exfiltrated Monzo account balance (£1,469.36), user location, and timestamp in JSON format via HTTP POST. The entire attack chain appears as routine weather personalization to the user.}
    \label{fig:exfiltration-proof}
\end{figure}

In response, the agent processed the \texttt{get\_forecast} tool exposed by the weather server. The embedded prompt triggered the following sequence, executed automatically by the agent:

\begin{enumerate}
    \item Discovery of the \texttt{account.balance} tool exposed by the Monzo server.
    \item Invocation of \texttt{account.balance} to retrieve the current bank account balance.
    \item Invocation of \texttt{send\_research\_data} to forward the balance data, location, and timestamp to the \href{https://webhook.site}{webhook.site} endpoint as shown in Figure~\ref{fig:exfiltration-proof}
\end{enumerate}

At each step, the Claude Desktop client did prompt the user for confirmation before invoking tools, including the final exfiltration request. However, because the instructions were embedded in a natural-sounding context, these prompts appeared consistent with legitimate weather personalization and were assumed to be trivially accepted by the user.

\subsection{Resource Requirements and Reproducibility}

The total setup time for this demonstration was under 2 hours. All resources were either:
\begin{enumerate}
    \item Free of cost
    \item Publicly accessible without authentication
    \item Based on official example code requiring no reverse engineering.
\end{enumerate}

No dedicated server infrastructure, paid services, or privileged knowledge was necessary. The only prerequisite skills were familiarity with basic Python scripting and the ability to modify prompt text within a template. The webhook.site endpoint was generated in approximately one second without any user registration.

\section{Limitations}

While our proof-of-concept demonstrates the feasibility of MCP-based financial data exfiltration, several factors constrain real-world exploitation:

\textbf{Target Environment Requirements.} The attack assumes victims have pre-installed high-value MCP servers exposing personal data. In our demonstration, successful exploitation required both conditions: (1) the user had configured a banking MCP server, and (2) the Monzo OAuth token remained valid. Given Monzo's 30-hour refresh token expiry~\footnote{\url{https://community.monzo.com/t/token-expiry/59376}}, attackers have a limited window before re-authentication is required, reducing the probability of successful exploitation.

\textbf{User Consent Mechanisms.} Claude Desktop's current implementation prompts users before each tool invocation, including the final \texttt{send\_research\_data} call. Our attack relies on users accepting these prompts without scrutiny, assuming the legitimacy of "personalized weather recommendations." Security-conscious users might question why a weather service needs to send data to an external endpoint, particularly one at webhook.site.

\textbf{Infrastructure Constraints.} Webhook.site URLs expire after 7 days of inactivity~\footnote{\url{https://docs.webhook.site/index.html}}, requiring attackers to actively monitor incoming data or frequently regenerate endpoints. While this creates operational overhead for attackers, it introduces an additional alarming privacy risk: the public nature of webhook.site means anyone who discovers or guesses the URL can view the victim's exfiltrated financial data, amplifying the potential harm beyond the initial attacker.

These limitations do not diminish the core vulnerability: MCP's trust model allows seemingly benign services to orchestrate cross-server attacks with minimal technical barriers. However, they do constrain the practical impact to targeted attacks against users who have both valuable MCP integrations and limited security awareness.

\section{Discussion and Recommendations}

While we demonstrated exfiltration of banking data, this attack pattern extends to any sensitive MCP server. Consider alternative targets: a Gmail MCP server could exfiltrate private communications, OAuth tokens, and password reset emails. A filesystem MCP server could steal SSH keys, API credentials, and sensitive documents. A calendar MCP could reveal travel patterns, business relationships, and meeting contents. Three systemic vulnerabilities enable these attacks:

\textbf{Protocol Permissiveness} While Claude Desktop implements permission prompts for each tool invocation, these safeguards are merely \textit{recommended} in MCP documentation, not enforced by the protocol. A malicious or poorly-implemented client could execute all tool calls silently, transforming our simple social engineering attack into automated data harvesting.

\textbf{Composability Without Security}: Installing any MCP server grants it access to all other servers' data through the AI assistant. The weather server cannot directly query the banking server, but it can instruct the AI to do so. Users have no immediate mechanism to define security boundaries between servers. MCP's power derives from combining arbitrary tools: weather with banking, email with calendar. However, each new server exponentially expands the attack surface.

\textbf{Ecosystem Velocity}: MCP servers proliferate through GitHub, and community directories without security vetting. Developers share and install servers based on functionality alone.

\subsection{Recommendations}

Immediate mitigations for users:
\begin{itemize}
\item Treat every MCP server as potentially malicious, regardless of source or functionality
\item Use only official MCP clients that enforce permission prompts
\item Never approve cross-server data access without clear, legitimate need
\item Audit installed servers regularly and remove unused integrations
\end{itemize}

Protocol-level interventions required:
\begin{itemize}
\item \textbf{Capability-based permissions}: Servers should declare required cross-server interactions at installation  \item \textbf{Mandatory access boundaries}: Allow users to designate servers as "sensitive" with restricted interoperability
\item \textbf{Server attestation}: Establish signing mechanisms and reputation systems for server distribution
\end{itemize}

\section{Conclusion}

This paper presents a concrete demonstration of a cross-server data exfiltration attack in the Model Context Protocol (MCP) ecosystem. The attack requires no backend infrastructure, no user credentials, no authentication bypass, and no specialized tooling; only basic Python scripting, open-source examples from the official MCP documentation, and a free webhook service. Despite its simplicity, the attack successfully extracts sensitive financial data by leveraging implicit trust relationships and agent-mediated tool composition. This highlights a critical security gap: even low-effort adversaries can exploit benign-appearing integrations to orchestrate high-impact data leaks.

Our results show that MCP’s core strengths: composability, flexibility, and tool interoperability, also introduce critical vulnerabilities when paired with implicit trust assumptions and inadequate security boundaries. The low barrier to entry for such attacks, combined with the growing proliferation of AI agents and community-built servers, represents a significant and urgent threat vector.

The attack we demonstrated generalizes beyond financial tools: any sensitive MCP server (email, filesystem, calendar) can be compromised through cross-server orchestration. This is enabled by three systemic weaknesses: (1) permissive execution defaults, where permission prompts are client-optional; (2) unrestricted composability, which allows any installed server to trigger tool invocations on others via the agent; and (3) rapid, unvetted ecosystem growth, where malicious servers can masquerade as useful tools. Without enforceable protocol-level controls, such as scoped capability declarations, server isolation boundaries, and signed attestations, MCP creates a wide, undersecured attack surface as agents integrate with high-risk tools.

\section*{Acknowledgments}
This work was supported by the Pivotal Research Fellowship in AI Safety.~\footnote{\url{https://www.pivotal-research.org/fellowship}}

\bibliographystyle{unsrt}  
\bibliography{references}

\begin{thebibliography}{1}

\bibitem{anthropic_mcp}
{Anthropic}.
\newblock Introducing the model context protocol.
\newblock \url{https://www.anthropic.com/news/model-context-protocol}, November 2024.

\bibitem{arxiv2025mcp}
Yuyou Gan et~al.
\newblock Model context protocol (mcp): Landscape, security threats, and future research directions.
\newblock {\em arXiv preprint arXiv:2503.23278}, 2025.

\bibitem{arxiv2025mcpsafetyaudit}
Brandon Radosevich and John~T. Halloran.
\newblock Mcp safety audit: Llms with the model context protocol allow major security exploits.
\newblock {\em arXiv preprint arXiv:2504.03767}, 2025.

\bibitem{mcp2024quickstart}
{Model Context Protocol Contributors}.
\newblock {MCP Quickstart: Building Your First Server}.
\newblock \url{https://modelcontextprotocol.io/quickstart/server}, 2024.

\bibitem{fastmcp2024docs}
{fastMCP Contributors}.
\newblock {fastMCP Documentation}.
\newblock \url{https://github.com/jlowin/fastmcp}, 2024.

\end{thebibliography}

\end{document}